\documentclass[11pt]{article}

\usepackage[dvips]{graphicx}
\usepackage{amssymb}
\usepackage{amsmath}
\usepackage{graphics}
\usepackage{graphicx}
\usepackage{xcolor}

\usepackage{amsfonts}

\newtheorem{theorem}{Theorem}[section]

\newtheorem{lemma}[theorem]{Lemma}

\newtheorem{proposition}[theorem]{Proposition}

\newtheorem{definition}[theorem]{Definition}

\usepackage{epsfig}
\usepackage{amsmath}
\usepackage{amsfonts}
\usepackage{amsfonts}

\newcommand{\nat}{{\mathbb N}}

\newcommand{\poly}{{\rm poly}}

\def\nottoobig#1{{\hbox{$\left#1\vcenter
to1.111\ht\strutbox{}\right.\n@space$}}}

\newcommand{\real}{\mathbb{R}}

\newcommand{\ie}{$\mbox{i.e.}$}

\newlength{\filength}
\newsavebox{\gcbox}
\sbox{\gcbox}{\framebox[\filength]{\rule{0ex}{2ex}}}

\newcommand{\qedblob}{\mbox{\rule[-1.5pt]{5pt}{10.5pt}}}
\def\literalqed{{\ \nolinebreak\hfill\mbox{\qedblob\quad}}}

\def\qed{\literalqed}

\newcommand{\singlespacing}{\let\CS=
\@currsize\renewcommand{\baselinestretch}{1}\tiny\CS}
\newcommand{\singlespacingplus}{\let\CS=
\@currsize\renewcommand{\baselinestretch}{1.25}\tiny\CS}
\newcommand{\doublespacing}{\let\CS=
\@currsize\renewcommand{\baselinestretch}{1.75}\tiny\CS}
\newcommand{\draftspacing}{\let\CS=
\@currsize\renewcommand{\baselinestretch}{2.0}\tiny\CS}

\def\zo{\{0,1\}}

\def\mapping{\rightarrow}

\newcommand{\dw}{{\rm DW}}
\newcommand{\w}{{\rm W}}
\newcommand{\ce}{${\rm{c.e.}}$}
\newcommand{\proof}{Proof.}
\newcommand{\rest}{{\upharpoonright}}
\newcommand{\prob}{{\rm Prob}}

\title{Possibilities and impossibilities in Kolmogorov complexity extraction}
\author{ {Marius Zimand\/}
\thanks{  Department of Computer and Information Sciences, Towson University,
Baltimore, MD.; email: mzimand@towson.edu; http://triton.towson.edu/\~{ }mzimand.
The author is supported in part
by NSF grant CCF 1016158.}}

\date{}
\begin{document}

\maketitle

\begin{abstract}
Randomness extraction is the process of constructing a source of randomness of high quality from one or several sources of randomness of lower quality. The problem can be modeled using probability distributions and min-entropy to measure their quality and also by using individual strings and Kolmogorov complexity to measure their quality. Complexity theorists are more familiar with the first approach. In this paper we discuss the second approach.  We present the connection between extractors and Kolmogorov extractors and the basic positive and negative results concerning Kolmogorov complexity extraction.
\end{abstract}

\section{Introduction}
Randomness is a powerful computational resource. For some problems, randomized algorithms are significantly faster than the best currently known deterministic algorithms. Furthermore, in some areas, such as cryptography, distributed computing, game theory, and machine learning, the use of randomness is compulsory, because some of the most basic operations simply do not have a deterministic implementation.
 
  It is not clear how to obtain the random bits that such algorithms need. While it seems that there are sources of genuine randomness in Nature, they produce sequences of bits with various biases and correlations that are not suitable for direct employment in some applications. For example in cryptographical protocols it is essential to use ``perfect'' or ``close-to-perfect'' randomness. 
  
  It thus is important to determine whether certain attributes of randomness can be improved effectively, or, even better, efficiently. It is obvious that randomness cannot be created from nothing ($\mbox{e.g.}$, from the empty string). On the other hand, it might be possible that if we already possess some randomness, we can produce ``better" randomness, or ``new" randomness.
  
  These general questions have been investigated in three settings:
  
  
\begin{itemize}
	\item[1] Finite probability distributions: We start with random variables $X_1$ over $\{0,1\}^{n_1}$,  $X_2$ over $\{0,1\}^{n_2}, \ldots$, $X_t$ over $\{0,1\}^{n_t}$, whose  distributions have min-entropy above a certain value that characterizes the quality of input randomness. We want a computable (or, better, a polynomial-time computable function) $f$ so that $f(X_1, \ldots, X_t)$ is close to the uniform distributions (that is $f$ produces ``better''  randomness), or $f(X_1, \ldots, X_t)$ is close to the uniform distributions even conditioned by some of the $X_i$'s (that is $f$ produces ``new'' randomness).
	
	\item[2] Finite binary strings: We start with finite binary strings $x_1 \in \zo^{n_1}$,  $x_2 \in \{0,1\}^{n_2}, \ldots$, $x_t \in \{0,1\}^{n_t}$, each string having some Kolmogorov complexity above a certain value that characterizes the quality of input randomness. We want a computable (or, better, a polynomial-time computable function) $f$ so that $f(x_1, \ldots, x_t)$ has close to maximum Kolmogorov complexity (that is $f$ produces ``better'' randomness), or $f(x_1, \ldots, x_t)$ has close to maximum Kolmogorov complexity even conditioned by some of the $x_i$'s (that is $f$ produces ``new'' randomness). 
	
		\item[3] Infinite binary sequences: We start with infinite binary sequences $x_1 \in \zo^{\infty}$,  $x_2 \in \{0,1\}^{\infty}, \ldots$, $x_t \in \{0,1\}^{\infty}$, each sequence having effective  Hausdorff dimension above a certain value that characterizes the quality of input randomness. We want a Turing reduction $f$ so that $f(x_1, \ldots, x_t)$ has effective Hausdorff dimension close to $1$  (that is $f$ produces ``better'' randomness), or $f(x_1, \ldots, x_t)$ has effective Hausdorff dimension close to $1$ even conditioned by some of the $x_i$'s (that is $f$ produces ``new'' randomness). 
\end{itemize}

The common scenario is that we start with $t$ sources (which are distributions, or strings, or sequences, depending on the setting), possessing some level of randomness, from which we want to obtain  better randomness and/or new randomness. This process is called randomness extraction.

Setting $1$ has been extensively studied and is familiar to the readers of this column. A function $f$ achieving the objective in setting $1$ is called an \emph{extractor}. Extractors have been instrumental in obtaining important results in derandomization, cryptography, data structures, and other areas.

In this paper we discuss \emph{Kolmogorov extractors}, which are the functions $f$ achieving the objectives in setting $2$ and settting $3$. The issue of Kolmogorov complexity extraction has been first raised for the case of infinite sequences by Reimann and Terwijn in 2003. The first explicit study for the case of finite strings is the paper by Fortnow, Hitchcock, A.~Pavan, Vinodchandran and Wang~\cite{fhpvw:c:extractKol} (versions of the problem have been investigated earlier, for example in~\cite{bfnv:c:increase-kol} and in~\cite{ver-vyu:j:kolm}). One reason for the late undertaking of this research line is the tight connection that exists between extractors and Kolmogorov extractors. However, Kolmogorov extractors have their own merits: they have applications in Kolmogorov complexity, algorithmic randomness, and other areas, and, perhaps more importantly, several general questions on randomness extraction, such as the amount of necessary non-uniformity   or the impact of input sources not being fully independent, are more natural to study in the framework of Kolmogorov extractors.

The paper is organized as follows. Section $2$ contains background information on Kolmogorov complexity. In Section $3$ we discuss Kolmogorov complexity extraction from finite strings (setting~$2$), and in Section $4$ we discuss Kolmogorov complexity extraction from infinite sequences (setting~ $3$). Section $5$ presents a few applications.
\section{Basic facts on Kolmogorov complexity}
The Kolmogorov complexity of a string $x$ is the length of the shortest effective description of $x$. There are several versions of this notion. We use here mainly the \emph{plain complexity}, denoted $C(x)$, and also the \emph{conditional plain complexity} of a string $x$ given a string $y$, denoted $C(x \mid y)$, which is the length of the shortest effective description of $x$ given $y$. The formal definitions are as follows.
We work over the binary alphabet $\zo$. A string is an element of $\{0,1\}^*$ and a sequence is an element of $\zo^{\infty}$.  
If $x$ is a string, $|x|$ denotes its length.  If $x$ is a sequence, then $x \rest n$ denotes the prefix of $x$ of length $n$.
Let $M$ be a Turing machine that takes two input strings and outputs one string. For any strings $x$ and $y$, define the \emph{Kolmogorov complexity} of $x$ conditioned by $y$ with respect to $M$, as 
$C_M(x \mid y) = \min \{ |p| \mid M(p,y) = x \}$.
There is a universal Turing machine $U$ with the following property: For every machine $M$ there is a constant $c_M$ such that for all $x$, $C_U(x \mid y) \leq C_M(x \mid y) + c_M$.
We fix such a universal machine $U$ and dropping the subscript, we write $C(x \mid y)$ instead of $C_U(x \mid y)$. We also write $C(x)$ instead of $C(x \mid \lambda)$ (where $\lambda$ is the empty string). The \emph{randomness rate} of a string $x$ is defined as ${\rm rate}(x) = \frac{C(x)}{|x|}$. 

In this paper, the constant hidden in the $O(\cdot)$ notation only depends on the universal Turing machine.

For all $n$ and $k \leq n$, $2^{k-O(1)} < |\{x \in \zo^n \mid C(x\mid n) < k\}| < 2^k$.


Strings $x_1, x_2, \ldots, x_k$ can be encoded in a self-delimiting way (\ie, an encoding from which each string can be retrieved) using $|x_1| + |x_2| + \ldots + |x_k| + 2 \log |x_2| + \ldots + 2 \log |x_k| + O(k)$ bits. For example, $x_1$ and $x_2$ can be encoded as $\overline{(bin (|x_2|)} 01 x_1 x_2$, where $bin(n)$ is the binary encoding of the natural number $n$ and, for a string $u = u_1 \ldots u_m$, $\overline{u}$ is the string $u_1 u_1 \ldots u_m u_m$ (\ie, the string $u$ with its bits doubled).

The Symmetry of Information Theorem (see \cite{zvo-lev:j:kol}) states that for all strings $x$ and $y$, $C(xy) \approx C(y) + C(y \mid x)$. More precisely:
$| (C(xy) - (C(x) + C(y \mid x)) | \leq O( \log C(x) + \log C(y))$.
In case the strings $x$ and $y$ have length $n$, it can be shown that
$| (C(xy) - (C(x) + C(y \mid x)) | \leq 2 \log n + O(\log \log n)$.

In Section~\ref{s:infinite}, we use a variant of Kolmogorov complexity, called \emph{prefix-free complexity} and denoted $K(x)$. The difference is that the underlying universal Turing machine $U$ is required to be a prefix-free machine, \ie, the domain of $U$ is a prefix-free set. It holds that for every string $x \in \zo^n$, $C(x) \leq K(x) \leq C(x) + O(\log n)$. Prefix-free sets over the binary alphabet have the following important property, called the \emph{Kraft-Chaitin} inequality. Let $\{n_1, n_2, \ldots, n_k, \ldots\}$ be a sequence of positive integers. Then $\sum 2^{-n_i} \leq 1$ iff there exists a prefix-free set $A = \{x_1, x_2, \ldots, x_k, \ldots \}$ with $|x_i| = n_i$, for all $i$. Moreover, if the sequence of lengths is computably enumerable (\ie, there is some computable $f$ such that $f(i) = n_i$ for all $i$) and the inequality holds, then $A$ is computably enumerable.

All the Kolmogorov extractors in this paper are ensembles of functions $f = (f_n)_{n \in \nat}$ of type $f_n : (\zo^n)^t \mapping \zo^{m(n)}$. The parameter $t$ is a constant and gives the number of input sources. In this survey we focus on the cases $t=1$ and $t=2$. Also note that we only consider the situation when all the sources have the same length. For readability, we usually drop the subscript and the expression ``ensemble $f: \zo^n \mapping \zo^m$'' is a substitute for
``ensemble $f = (f_n)_{n\in \nat}$, where for every $n$, $f_n : \zo^n \mapping \zo^{m(n)}$, and similarly for the case of more sources.

\section{The finite case}
\label{s:finite}
\subsection{Kolmogorov extraction from one string}
We first consider Kolmogorov extraction when the source consists of a single binary string $x$ that contains some complexity. For concreteness, think of the case when $C(x) \geq \sigma n$, where $n = |x|$ and $\sigma$ is a positive constant. If $\sigma$ is the only information that the extractor has about the source, then Kolmogorov extraction is impossible, as one can see from the following simple observation.
\begin{proposition}
\label{p:basic}
Let $f: \zo^n \mapping \zo^m$ be a uniformly computable ensemble of functions. Then, for every $n$, there exists a string $x$ in $\zo^n$ with $C(x) \geq n-m$ and $C(f(x) \mid n) = O(1)$.
\end{proposition}
$\proof$ Let $z$ be the most popular string in Image($f(\zo^n)$) (\ie, with the largest number of preimages), with ties broken in some canonical way. Since the above is a full description of $z$, $C(z \mid n) = O(1)$. The string $z$ has at least $2^{n-m}$ preimages and, therefore, there exists a string $x$ in the preimage set of $z$ with $C(x) \geq n-m$.~\qed
\smallskip

In particular, if $m \leq \sigma n$ and $\sigma \leq 1/2$, there exists a string $x \in \zo^n$ with $C(x) \geq \sigma n$ and $C(f(x) \mid n) = O(1)$.

Kolmogorov extraction may be possible if the extractor possesses additional information about the source $x$. We call this \emph{advice about the source}. The basic case is when the extractor knows $C(x)$.  Then, one can construct $x^*$, a shortest description of $x$. In other words,  $C(x^* \mid x) \leq \log C(x) + O(1) \leq \log n + O(1)$, and it is easy to see that $C(x^*) \geq |x^*| - O(1)$. Thus, with at most $\log n + O(1)$ bits of advice about the source $x$, one can essentially extract all the randomness in the source. Buhrman, Fortnow, Newman and Vereshchagin~\cite{bfnv:c:increase-kol} have shown how to extract in polynomial time almost all the randomness in the source with $O(\log n)$ advice about the source.

Fortnow et al.~\cite{fhpvw:c:extractKol} have shown that with a \emph{constant} number of advice bits about the source, one can increase the randomness rate to arbitrarily close to $1$. Moreover, their Kolmogorov extractor runs in polynomial time. 
\begin{theorem}[\cite{fhpvw:c:extractKol}]
\label{t:fortnowKolmextract}
For any rational $\sigma > 0$, $\epsilon > 0$, there exists a polynomial-time computable function $f$ and a constant $k$ such that for any $x$ with ${\rm rate}(x) \geq \sigma$, it holds that ${\rm rate}(f(x, \alpha_x)) \geq 1 - \epsilon$ for some string $\alpha_x$ of length $k$. The length of $f(x, \alpha_x)$ is at least $C |x|$, for a constant $C$ that only depends on $\sigma$ and $\epsilon$.
\end{theorem}
A sketch of the proof is given in Section~\ref{s:twosources}, after we present the relation between extractors and Kolmogorov extractors. In the opposite direction,
Vereshchagin and Vyugin~\cite{ver-vyu:j:kolm} show the limitations of what can be extracted with a bounded quantity of advice. To state their result, let us fix $n = $ length of the source, $k =$ number of bits of advice that is allowed, and $m = $ the number of extracted bits. Let  $K = 2^{k+1}-1$.
\begin{theorem}[\cite{ver-vyu:j:kolm}]
\label{t:vervyu}
There exists a string $x \in \zo^n$  with $C(x) > n- K \log (2^m+1) \approx n - Km$ such that any string $z \in \zo^m$ with $C(z \mid x) \leq k$ has complexity $C(z) < k + \log n + \log m + O(\log \log n, \log \log m)$. 

In other words, any string $z$ that is effectively obtained from $x$ with $k$ bits of advice, has in fact unconditional complexity $\approx k$.
\end{theorem}
$\proof$ For each $x \in \zo^n$, let ${\rm Range}(x) = \{z \in \zo^m \mid C(z \mid x) \leq k\}$.  Similarly to the proof of Proposition~\ref{p:basic}, the idea is to produce a set of strings in $\zo^m$ that is ``popular," in the sense that is equal to ${\rm Range}(x)$, for many $x \in \zo^n$ (we refer to these sets as Ranges). Let $T = 2^{m}+1$. In a dovetailing manner, we run $U(p,x)$ for all $x \in \zo^n$, and all $p \in \zo^{\leq k}$. We call this an enumeration procedure. Note that if $U(p,x)$ halts, it outputs a string in ${\rm Range}(x)$.  In step 1, we run this enumeration till it produces a string $z_1$ that belongs to at least $2^n/T$ Ranges. There may be no such $z_1$ and we deal with this situation later. We mark with (1) all these Ranges. In step $2$, we resume the enumeration procedure till it produces a string $z_2$ different from $z_1$ that belongs to at least a fraction $1/T$ of the Ranges marked (1). We re-mark this ranges with (2). In general, at step $i$, we run the enumeration till it produces a string 
$z_i$ that is different from the already produced strings and that belongs to at least a fraction of $1/T$ of the Ranges marked $(i-1)$ at the previous step. We re-mark these Ranges with $(i)$. We continue this operation till either (a) we have completed $K$ steps and have produced $K$ strings $z_1, \ldots, z_K \in \zo^m$, or (b) at some step $i$, the enumeration fails to find $z_i$. In case (a), there are at least $2^n/T^K$ Ranges that are equal to $\{z_1, \ldots, z_K\}$. In case (b), there are $2^n/T^{i-1}$ that have $\{z_1, \ldots, z_{i-1}\}$ as a subset. In addition, for every $z \in \zo^m - \{z_1, \ldots, z_{i-1}\}$, the set $\{z_1, \ldots, z_{i-1}, z\}$ is a subset of less than $2^n/T^i$ ranges. It means that $\{z_1, \ldots, z_{i-1}\}$ is equal to at least $2^n/T^{i-1} - 2^m \cdot 2^n/T^i = 2^n/T^i$ Ranges. Consequently, the procedure produces a set $\{z_1, z_2, \ldots, z_s\}$, $s \leq K$, that is equal to ${\rm Range}(x)$ for at least $\frac{2^n}{T^K} = \frac{2^n}{(1+2^m)^K}$  strings $x \in \zo^n$. One of these strings $x$ must have Kolmogorov complexity $C(x) \geq n - K \log(2^m+1)$. Each string $z_i$ produced by the procedure can be described by $i \leq K$, by $n$, by $m$, and by $k$. We represent $i$ on exactly $k+1$ bits, and this will also describe $k$. Thus, $C(z_i) \leq k + \log n + \log m + O(\log \log n, \log \log m)$.~\qed
\smallskip

Vereshchagin and Vyugin's result explains why the Kolmogorov extractor in Theorem~\ref{t:fortnowKolmextract} does not achieve rate $1$. Theorem~\ref{t:vervyu} implies that if a single-source Kolmogorov extractor increases the rate from $\sigma$ to $1-\epsilon$ using $k$ bits of advice, then $\epsilon = \Omega\big(\frac{1-\sigma}{2^{k} }\big)$ (provided  that the output length $m$ is a constant fraction of $n$).
\subsection{Kolmogorov extraction from two strings}
\label{s:twosources}
We recall that a Kolmogorov extractor with two sources is an ensemble of functions of the type $f: \zo^n \times \zo^n \mapping \zo^m$. The quality of the two sources is given by their Kolmogorov complexity and by their degree of dependency. The dependency of two strings is the amount of information one string has about the other string.
\begin{definition}[Dependency]
\label{d:dep}
For $x \in \zo^n$, $y \in \zo^n$, the dependency of $x$ and $y$ is given by
\[
{\rm dep}(x,y) = \max\{C(x \mid n) - C(x \mid y), C(y \mid n) - C(y \mid x)\}.
\]
\end{definition}
There are in the literature several variations of the above definition. They all differ by at most an $O(\log n)$ additive term. For example, one may prefer $C(x) + C(y) - C(xy)$ as a value that captures the dependency of $x$ and $y$. It holds that $|(C(x) + C(y) - C(xy)) - {\rm dep}(x,y)| = O(\log n)$. Definition~\ref{d:dep} tends to produce sharper statements.

The class of sources from which we extract is characterized by two parameters: $k$ = the minimum Kolmogorov complexity that each input string has, and $\alpha$ = the maximum dependency of the input strings. Accordingly, for positive integers $k$ and $\alpha$, we let
\[
S_{k,\alpha} = \{(x,y) \in \zo^n \times \zo^n \mid C(x\mid n) \geq k, C(y \mid n) \geq k, {\rm dep}(x,y) \leq \alpha\}.
\]
In other words, $S_{k,\alpha}$ consists of those pairs of input sources that have complexity at least $k$ and dependency at most $\alpha$.
\begin{definition}[Kolmogorov extractor] An ensemble of functions $f: \zo^n \times \zo^n \mapping \zo^m$ is a $(k,\alpha, d)$ Kolmogorov extractor if for every $(x,y) \in S_{k,\alpha}$,
$C(f(x,y)\mid n) \geq m - d$.
\end{definition}
\subsubsection{The curse of dependency: limitations of Kolmogorov extractors with two sources}
\label{s:stringimpossible}
As we have discussed above, we would like to have a computable function $f: \zo^n \times \zo^n \mapping \zo^m$ such that for all $(x,y) \in S_{k, \alpha}$, $C(f(x,y)) \approx m$.  For a string $z$, we define its \emph{randomness deficiency} to be  $|z| - C(z)$ and thus we would like the randomness deficiency of $f(x,y)$ to be $\approx 0$. However, we will see that this is impossible. We observe that no computable function $f$ as above can guarantee that for all $(x,y) \in S_{k,\alpha}$ the randomness deficiency of $f(x,y)$ is less than $\alpha - O(\log \alpha) + O(1)$, even for a large value of $k$.

\begin{theorem}[\cite{zim:c:impossibamplific}]
\label{t:cursedep}
%
There is no uniformly computable ensemble of functions  $f : \zo^n \times \zo^n \mapping \zo^m$, such that 
for all $(x,y) \in S_{k,\alpha}$, the randomness deficiency$(f(x,y)) \leq \alpha - O(\log \alpha)$.

The above holds for all $k \leq n - \alpha$ and all $m \geq \alpha$ (ignoring $O(\log n)$ additive terms).
\end{theorem}
$\proof$
Let $f : \zo^n \times \zo^n \mapping \zo^m$ be a uniformly computable ensemble of functions.
We look at prefixes of length $\alpha$ of strings in the image of $f$.
	 Let $z$ be the most popular prefix of length $\alpha$ of strings in the image of $f$.
Note that	 $C(z\mid n) = O(1)$.
	There are $\geq 2^{2n - \alpha}$ pairs $(x,y)$ with $f(x,y) \rest \alpha = z$.
	 There is a pair $(x,y)$ as above with $C(xy \mid n) \geq 2n - \alpha$.
 It follows that $(x,y) \in S_{n-\alpha, \alpha}$ (ignoring $O(\log n)$ terms).
	 Since $f(x,y) = zw$ with $|w|= m-\alpha$, it follows that $C(f(x,y)|n) \leq m - \alpha + 2 \log \alpha +O(1)$.
	In other words, the randomness deficiency of $f(x,y)$ is at least $\alpha - 2 \log \alpha - O(1)$.~\qed
\subsubsection{Extractors {vs.} Kolmogorov extractors} 
Positive results (within the limitations shown in Theorem~\ref{t:cursedep}) regarding Kolmogorov complexity extraction can be obtained by exploiting the relation between extractors and Kolmogorov extractors.

In the case of extractors, sources are modeled by random variables taking values over $\zo^n$. Sometimes, such a random variable is indentified with the distribution of its output. 
The min-entropy of a distribution $X$ over 
$\zo^n$, denoted $H_{\infty}(X)$, is given by
$H_{\infty}(X) = \min \Big\{ \log \frac{1}{\prob(X=a)}  \mid a \in \zo^n, \prob(X=a) \not= 0 \Big\}$.
Thus if $X$ has min-entropy $\geq k$, then for all $a$ in the range of $X$, ${\prob(X=a) \leq 1/2^k}$.  
A distribution $X$ over $\zo^n$ with min-entropy $k$ is called an $(n,k)$-source.
For each $n \in \nat$, let $U_n$ denote the uniform distribution over $\zo^n$. 
The min-entropy of a source is a good indicator of the quality of its randomness. Note that if $H_{\infty}(X) = n$, then $X = U_n$, and thus $X$ is ``perfectly'' random. Smaller values of min-entropy indicate defective sources (the smaller the min-entropy is, the more defective the source is).

For $A \subseteq \zo^n$, we denote $\mu_X(A) = \prob(X \in A)$. Let $X$, $Y$ be two sources over~$\zo^n$. 
The distance between two distributions $X$ and $Y$ over $\zo^n$ is $|X-Y| = max_{A \subseteq \zo^n}|\mu_X(A) - \mu_Y(A)|$. It is easy to show that
$|X-Y|= (1/2) \sum_{a \in \zo^n} |\mu_X(a) - \mu_Y(a)| = \sum_{a: \mu_X(a) \geq \mu_Y(a)} \mu_X(a) - \mu_Y(a)$. The distributions $X$ and $Y$ are $\epsilon$-close if $|X-Y| \leq \epsilon$. The following facts are helpful.
\begin{lemma} [\cite{guv:j:extractor}]
\label{l:closedistributions}
Let $D$ be a distribution over $\zo^n$ and let ${\rm HEAVY}_{k,t} = \{a \in \zo^n \mid \mu_D(a) > t  2^{-k}\}$. 

(1) If $D$ is $\epsilon$-close to a distribution with min-entropy $k$ then $\mu_D({\rm HEAVY}_{k,t}) \leq 1/t+\epsilon$.

(2) Suppose that for every set $S \subseteq \zo^n$ of size $K$, $\mu_D(S) \leq \epsilon$. Then $D$ is $\epsilon$-close to a distribution with min-entropy at least $\log (K/\epsilon)$.
\end{lemma}
$\proof$ (1)  Let $D'$ be a distribution with min-entropy $k$ such that $|D- D'| \leq \epsilon$. 
Since $1 \geq \mu_D({\rm HEAVY_{k,t}}) \geq |{\rm HEAVY_{k,t}}|\cdot t2^{-k}$, we have $|{\rm HEAVY_{k,t}}| \leq 2^k/t$. Then $\mu_{D'}({\rm HEAVY_{k,t}}) \leq |{\rm HEAVY_{k,t}}|\cdot 2^{-k} \leq 1/t$ and, therefore, $\mu_{D}({\rm HEAVY_{k,t}}) \leq \mu_{D'}({\rm HEAVY_{k,t}}) + \epsilon \leq 1/t + \epsilon$.~\qed

(2) Let $x_1, x_2, \ldots, x_N$ be an ordering of $\zo^n$ such that $\mu_D(x_1) \geq \mu_D(x_2) \geq \ldots \geq \mu_D(x_N)$. Let $2^{-\ell} = (\mu_D(x_1) + \mu_D(x_2) + \ldots + \mu_D(x_K))/K$ (the average of the heaviest $K$ elements). Note that each of the elements $x_{K+1}, \ldots, X_N$ has mass at most $2^{-l}$. Also, since $\mu_D(x_1) + \ldots + \mu_D(x_K) \leq \epsilon$, we have $\ell \geq \log(K/\epsilon)$. Consider the distribution $D'$ that assigns mass $2^{-\ell}$ to each of $x_1, \ldots, x_K$ and is the same as $D$ on the elements $X_{K+1}, \ldots, x_N$. Then $D'$ has min-entropy $\ell \geq \log(K/\epsilon)$ and $|D - D'| = \sum_{a : \mu_D(a) \geq \mu_{D'}(a)} \mu_D(a) - \mu_{D'}(a) \leq \mu_D(\{x_1, \ldots, x_K\}) \leq \epsilon$.~\qed

It turns out that Kolmogorov extractors are roughly equivalent to \emph{almost extractors}, which are in general weaker than extractors (two-source extractors is what we obtain if we take $d=0$ in the next definition).
\begin{definition}[Almost extractor]
An ensemble of functions $f: \zo^n \times \zo^n \mapping \zo^m$ is a $(k, \epsilon, d)$ almost extractor if for all independent random variables $X$ and $Y$ over $\zo^n$ with $H_\infty(X) \geq k$ and $H_\infty(Y) \geq k$, the random variable $f(X,Y)$ over $\zo^m$ is $\epsilon$-close to a distribution $D$ on $\zo^m$ having $H_\infty(D) \geq m-d$.
\end{definition} 
A very useful result of Chor and Goldreich~\cite{cho-gol:j:weaksource} states that, in the above definition, it is enough to restrict the requirement to all random variables having a flat distribution, \ie, a distribution that assigns equally the probability mass to the elements of a set of size $2^k$.

The connection between almost extractors and Kolmogorov extractors is most easily understood by looking at their combinatorial characterizations. The relevant combinatorial object is that of a \emph{balanced table}. The approach is to view a function $f: \zo^n \times \zo^n \mapping \zo^m$ as a table with rows in $[N]$, columns in $[N]$, and colored with colors from $[M]$, where $N=2^n, M=2^m$ and we identify $\zo^n$ with $[N]$ and $\zo^m$ with $[M]$. For a set of colors $U \subseteq [M]$, an $U$-cell is a cell in the table whose color is in $U$. A rectangle is the restriction of $f$ to a set of the form $B_1 \times B_2$, where $B_1 \subseteq [N], B_2 \subseteq [N]$. The balancing property requires that in all  rectangles of size $2^k$-by-$2^k$, all colors appear approximately the same number of times. Depending on how we quantify ``approximately," we obtain different types of balanced tables. Also, sometimes, we require the balancing property to hold not for every individual color $a \in [M]$, but for each set of colors $U \subseteq [M]$ of a given size. 

To get an intuition on why balanced tables are relevant for randomness extraction, it is easier to consider the case of Kolmogorov extractors. To make matters concrete, suppose we are shooting for a Kolmogorov extractor with complexity parameter $k$ and dependency parameter $\alpha$.
If the $[N]$-by-$[N]$ table $f$ colored with colors in $[M]$ is not balanced, then there is an element in the range of $f$ that has many preimages. Arguing as in the proof of Proposition~\ref{p:basic}, this implies that $f$ is not a Kolmogorov extractor. For the other direction, let us fix $(x,y) \in S_{k, \alpha}$. Let $B_x = \{u \in \zo^n \mid C(u \mid n) \leq C(x \mid n)\}$ and $B_y = \{v \in \zo^n \mid C(v \mid n) \leq C(y \mid n)\}$. $B_x \times B_y$ forms a rectangle of size $\approx 2^{C(x|n)} \times 2^{C(y\mid n)}$, and this is $\approx 2^k \times 2^k$ or larger (because $C(x \mid n) \geq k, C(y \mid n) \geq k$). Suppose that the table $f$ satisfies the following balancing property: Each color from $[M]$ appears in the rectangle $B_x \times B_y$ a fraction of at most $c/M$ times, where $c$ is a constant. Clearly, $(x,y)$ is a cell in $B_x \times B_y$ and, therefore, the color $z = f(x,y)$ appears at most $(c/M) \cdot 2^{C(x\mid n) + C(y\mid n)} = 2^{C(x\mid n) + C(y \mid n)- m+O(1)}$ times in $B_x \times B_y$. If $C(x \mid n)$ and $C(y \mid n)$ are given, one can effectively enumerate the elements of $B_x \times B_y$. Then the string $xy$ can be described by $z$, by $C(x \mid n)$ and $C(y \mid n)$, by the rank $r$ of the cell $(x,y)$ in an enumeration of the $z$-colored cells in $B_x \times B_y$, and by the table $f$. Thus, $C(xy \mid n) \leq C(z \mid n) + C(C(x \mid n)) + C(C(y \mid n)) + C(r \mid n) + C(\mbox{table} \mid n) + O(\log n)$. $C(C(x \mid n))$ and $C(C(y\mid n))$ are $O(\log n)$, and, since  the table is computed from $n$ (because the ensemble $f$ is uniformly computable), $C(\mbox{table}\mid n) = O(1)$. By the above estimation, $C(r \mid n) \leq C(x \mid n) + C(y \mid n) - m +O(1)$. We obtain $C(xy \mid n) \leq C(z \mid n) + C(x \mid n) + C(y \mid n) - m + O(\log n)$.
 On the  other hand, from the dependency property of $x$ and $y$, $C(xy \mid n) \geq C(x\mid n) + C(y \mid n) - \alpha$. It follows that $C(z \mid n) \geq m - \alpha - O(\log n)$, which is the desired conclusion. With a more elaborate argument, we can get $O(1)$ instead of $O(\log n)$. Since we need the above to be true for every $(x,y) \in S_{k,\alpha}$, we require that the above balancing property holds for all rectangles of size $2^k \times 2^k$, or larger. In fact it is enough to require the balancing property to hold for all rectangles of size $2^k \times 2^k$ (because if there exists a larger unbalanced rectangle, then there is also a $2^k \times 2^k$ unbalanced rectangle).
 
 After this motivating discussion, we pursue with the combinatorial characterization of almost extractors and of Kolmogorov extractors.
\begin{proposition}[Combinatorial characterization of almost extractors]
\label{p:combextractor}
Let $f: \zo^n \times \zo^n \mapping \zo^m$ be an ensemble of functions.

(1) If $f$ is a $(k, \epsilon, d)$ almost extractor, then for every rectangle $B_1 \times B_2 \subseteq [N] \times [N]$ of size $2^k \times 2^k$, and  for any set of colors $U \subseteq [M]$,
\[
\frac{|\{U\mbox{-cells in } B_1 \times B_2\}|}{|B_1 \times B_2|} \leq \frac{|U|}{M}\cdot 2^{d} + \epsilon.
\]
(2) Suppose that for every rectangle $B_1 \times B_2 \subseteq [N] \times [N]$ of size $2^k \times 2^k$, for any set of colors $U \subseteq [M]$ with $|U| = \epsilon \cdot M \cdot 2^{-d}$,
\[
\frac{|\{U\mbox{-cells in } B_1 \times B_2\}|}{|B_1 \times B_2|} \leq \frac{|U|}{M}\cdot 2^{d} + \epsilon.
\]
Then $f$ is a $(k, 2\epsilon, d)$ almost extractor.
\end{proposition}
$\proof$ (1). Let $X$ and $Y$ be two independent random variables that are flat on $B_1$, respectively $B_2$. Since $X$ and $Y$ have min-entropy $k$, $f(X,Y)$ is $\epsilon$-close to a distribution $D$ with min-entropy at least $m-d$. We have $\mu_D(U) \leq |U| \cdot 2^{-m+d}$ and the conclusion follows because $\mu_{f(X,Y)}(U) \leq \mu_D(U) + \epsilon$ and
$\mu_{f(X,Y)}(U) = \frac{|\{U\mbox{-cells in } B_1 \times B_2 |\}}{|B_1 \times B_2|}$.~\qed

(2) Let $X$ and $Y$ be independent random variables that have flat distributions over $\zo^n$ with min-entropy $k$. Let $B_1$ be the support of $X$ and $B_2$ be the support of $Y$. Then $\mu_{f(X,Y)}(U) = \frac{|\{U\mbox{-cells in } B_1 \times B_2\}|}{|B_1 \times B_2|} \leq \frac{|U|}{M}\cdot 2^{d} + \epsilon \leq 2 \epsilon$ (the first equality holds because $X$ and $Y$ are flat, and the second and third inequalities follow from the hypothesis). Then, by Lemma~\ref{l:closedistributions} (2), $f(X,Y)$ is $2\epsilon$-close to a distribution with min-entropy equal to
$\log(|U|/\epsilon) = m-d$.~\qed
\smallskip

\begin{proposition}[Combinatorial characterization of Kolmogorov extractors]
\label{p:combKolm}
Let $f:\zo^n \times \zo^n \mapping \zo^m$ be an ensemble of functions.

(1) If $f$ is a $(k, \alpha, d)$-Kolmogorov extractor, then for any rectangle $B_1 \times B_2 \subseteq [N] \times [N]$ of size $2^{k'} \times 2^{k'}$, where $k'= k+\alpha$, for any set of colors $U \subseteq [M]$, with size $|U| = 2^{-\alpha}\cdot M \cdot 2^{-(d+O(1))}$, it holds that 
\[
\frac{|\{U\mbox{-cells in } B_1 \times B_2\}|}{|B_1 \times B_2|} \leq \frac{|U|}{M}\cdot 2^{d+O(1)}.
\]

(2) Suppose that there exists a constant $d$ such that for all rectangles $B_1 \times B_2$ of size $2^k \times 2^k$, for any $U \subseteq [M]$ and for some $\epsilon$ computable from $n$, it holds that  
\[\frac{|\{U\mbox{-cells in } B_1 \times B_2\}|}{|B_1 \times B_2|} \leq \frac{|U|}{M}\cdot 2^{d} + \epsilon.
\]
Then $f$ is a $(k', \alpha, \alpha + 2d+1)$ Kolmogorov extractor, where $k'= k+ \log n +O(\log \log n)$, and $\alpha = \log (1/\epsilon) + d + 1$.

\end{proposition}
$\proof$ (1) Suppose there exist $B_1, B_2, U$ violating the conclusion. Specifically we assume: $B_1 \subseteq [N], |B_1| = 2^{k'}$, $B_2 \subseteq [N], |B_2| = 2^{k'}$, where $k'=k+\alpha$, $U\subseteq [M]$, $|U| = 2^{m-\alpha -d-c_1}$ and $\frac{|U\mbox{-cells in } B_1 \times B_2|}{|B_1 \times B_2|} > 4 \cdot 2^{- \alpha +c_1}$, for appropriate choices of the constants.  We construct the first (in some canonical sense) triplet $(B_1, B_2, U)$ satisfying the above relations. Note that for every $z \in U$, $C(z \mid n) \leq m - \alpha - d - c_1 +O(1) < m-\alpha -d$, for a sufficiently large $c_1$.  

We estimate the number of elements of $B_1 \times B_2$ that are not good for extraction, \ie, the size of $B_1 \times B_2 - \overline{S_{k, \alpha}}$. $B_1 \times B_2 - \overline{S_{k, \alpha}}$  is contained in the union of ${\rm BAD}_1$, ${\rm BAD}_2$, and ${\rm BAD}_3$, where
${\rm BAD}_1 = \{(x,y) \in  B_1 \times B_2 \mid C(x \mid n) < k \}$, ${\rm BAD}_2 = \{(x,y) \in  B_1 \times B_2 \mid C(y \mid n) < k \}$ and
${\rm BAD}_3 = \{(x,y) \in  B_1 \times B_2 \mid C(x \mid n) - C(x\mid y) > \alpha \mbox{ or } C(y \mid n) - C(y\mid x) > \alpha \}$.
Clearly, $|{\rm BAD}_1|$ and $|{\rm BAD}_2|$ are each bounded by $2^{k+k'}$. Regarding ${\rm BAD}_3$, note that if $C(x \mid n) - C(x \mid y) > \alpha$, then $C(x \mid y ) < C(x \mid n) - \alpha < k' - \alpha +O(1)$ (because, conditioned by $n$, $x$ can be described by its rank in a canonical enumeration of $B_1$). Similarly, if $C(y \mid n) - C(y \mid x) > \alpha$, then $C(y \mid x )  < k' - \alpha +O(1)$.  It follows that $|{\rm BAD}_3| \leq 2 \cdot 2^{2k' - \alpha +O(1)}$. Thus, $|B_1 \times B_2 - \overline{S_{k, \alpha}}| \leq  |{\rm BAD}_1| + |{\rm BAD}_2|+ |{\rm BAD}_3| \leq 2^{k+k'} + 2^{k+k'} + 2 \cdot 2^{2k' - \alpha +O(1)} \leq 4 \cdot 2^{2k' - \alpha + c_1}$, for a sufficiently large $c_1$. Since the number of $U$-cells in $B_1 \times B_2 > 4 \cdot 2^{2k' - \alpha + c_1}$, there exists a pair $(x,y) \in B_1 \times B_2 \cap S_{k,\alpha}$ such that $f(x,y) \in U$. Let $z = f(x,y)$. It follows that $z \in U$ and $C(z \mid n) \geq m - \alpha - d$, contradiction.~\qed
\smallskip

(2) Fix $(x,y) \in S_{k',\alpha}$. Let $z = f(x,y)$ and let $t=\alpha + 2d + 1$. For the sake of contradiction, suppose that $C(z\mid n) < m-t$. Let $t_x =C(x \mid n) \geq k'$ and $t_y = C(y \mid n) \geq k'$. Let $B_x = \{u\in \zo^n \mid C(u \mid n) \leq t_x\}$ and $B_y = \{v\in \zo^n \mid C(v \mid n) \leq t_y\}$. Note that
$2^{t_x - O(1)} \leq |B_x| \leq 2^{t_x+1}$ and $2^{t_y - O(1)} \leq |B_y| \leq 2^{t_y+1}$. We  take $U = \{u \in \zo^m \mid C(u \mid n) < m-t\}$.
We have $|U|/M \cdot 2^d + \epsilon \leq (2^{m-\alpha - 2d-1}/2^m) \cdot 2^d + 2^{-\alpha - d -1} = 2^{-\alpha - d}$. We say that a column $v \in [N]$ is \emph{bad} if the number of $U$-cells in $B_x \times \{v\}$ is $\geq 2^{t_x - \alpha -d}$. The number of bad columns is $< 2^k$ (otherwise the hypothesis would be violated by the rectangle formed with $B_x$ and the set of bad columns). Also, the set of bad columns can be enumerated if $n$ and $t_x$ are given. It follows that if $v$ is a bad column, then
$C(v \mid n) < k + \log t_x + 2 \log \log t_x +O(1) < k + 2 \log n$. Since $C(y \mid n) \geq k'$, $y$ is a good column. Therefore, the
number of $U$-cells in $B_x \times \{y\}$ is $< 2^{t_x - \alpha - d}$. By our assumption, $(x,y)$ is an $U$-cell in $B_x \times \{y\}$. So, the string $x$ can be described by: $y$, rank of $(x,y)$ in an enumeration of $U$-cells in  $B_x \times \{y\}$, $t_x$ and $d$. We write the rank on exactly $t_x - \alpha - d$ bits and this also provides $t_x$. It follows that $C(x \mid y) \leq t_x - \alpha - d + \log d + 2 \log \log d +O(1)$. On the other hand, since ${\rm dep}(x,y) \leq \alpha$, $C(x \mid y) \geq t_x - \alpha$. It follows that $d \leq \log d + 2 \log \log d + O(1)$, contradiction (if $d$ is large enough).~\qed

Combining the combinatorial characterizations of almost extractors and of Kolmogorov extractors, we obtain the following theorem.
\begin{theorem}[Equivalence of almost extractors and Kolmogorov extractors]
\label{t:ExtKolmExt}
Let $f: \zo^n \times \zo^n \mapping \zo^m$ be an ensemble of functions.

(1) (implicit in~\cite{fhpvw:c:extractKol}) If $f$ is a $(k, \epsilon,d)$ almost extractor, then $f$ is a $(k', \alpha, \alpha +2d+1)$ Kolmogorov extractor, where $k'=k+\log n + O(\log \log n)$ and $\alpha = \log(1/\epsilon) + d +1$.

(2) (\cite{hit-pav-vin:c:Kolmextraction}) If $f$ is a $(k, \alpha, d)$ Kolmogorov extractor, then $f$ is a $(k', \epsilon, d')$ almost extractor, where $k' = k+ \alpha$, $\epsilon = 2 \cdot 2^{-\alpha}$ and $d' = d + O(1)$. 
\end{theorem}
In brief, any almost extractor is a Kolmogorov extractor with a small increase in the min-entropy parameter, and vice-versa. In the correspondence between the two notions, the dependency parameter of the Kolmogorov extractor and the error parameter of the almost extractor are related by $\alpha \approx \log 1/\epsilon$.

Thus, we can take any two-source extractor (recall that any two-source extractor is an almost extractor with the randomness deficieny parameter $d=0$), and immediately conclude that it is also a Kolmogorov extractor. Dodis and Oliveira~\cite{dod-oli:c:extractor} showed the existence of  computable $(k,\epsilon)$ two-source extractors for any $k \geq \log n + 2 \log 1/\epsilon$, with output length $m= 2k - 2 \log 1/\epsilon$. In applications, we typically need polynomial-time computable procedures. If we focus on the min-entropy parameter, the currently best polynomial-time two-source extractors are due to Bourgain~\cite{bou:j:multiextract}, which has $k = 0.4999n$ and $m = \Omega(n)$, and to Raz~\cite{raz:c:multiextract}, in which one source needs to have min-entropy $> 0.5n$ and, the second one only need to have min-entropy polylog($n$). Kalai, Li, and Rao~\cite{kal-li-rao:c:condextractor} have used a hardness assumption to construct a polynomial-time two-source extractor for min-entropy $\delta n$ (for both sources, and constant $\delta$) and $m = n^{\Omega(1)}$. The hardness assumption is the existence of one-way permutations with certain parameters.  For sources with min-entropy $ > 0.5 n$, Shaltiel~\cite{sha:c:longerextract}, has constructed a polynomial-time two-source extractor 
with $k = (1/2 + \delta n)$, $\epsilon = 2^{-\log ^4 n}$, and $m = 2k- c \log(1/\epsilon)$, where $c$ is a constant that depends on $\delta$. Rao~\cite{rao:c:randpm} has constructed a polynomial-time computable $(k, \epsilon, d)$ almost extractor for $k = \delta n$, $d = \poly(1/\delta, 1/\epsilon)$ and $m = O(\delta n)$. By Theorem~\ref{t:ExtKolmExt}, all these results lead to Kolmogorov extractors with the corresponding parameters.

Radhakrishnan and Ta-Shma~\cite{rad-tas:j:extractors} showed that any two-source extractor must suffer an entropy loss of $2 \log 1/\epsilon$.  Thus, any two-source extractor $E: \zo^n \times \zo^n \mapping \zo^m$, with parameters $(k, \epsilon)$, must have output length $m \leq 2k - 2 \log 1/\epsilon$. When we view $E$ as a $(k + O(\log n), \alpha, d)$ Kolmogorov extractor, via Theorem~\ref{t:ExtKolmExt}, the dependency parameter $\alpha$ is $\approx \log 1/\epsilon$. Recall that the randomness deficiency of a Kolmogorov extractor is at least $\alpha$, which, in other words, means that $C(E(x,y)) \leq m - \alpha$. Thus, at best, we obtain that for any $(x,y) \in S_{k,\alpha}$,
$C(E(x,y)) \approx 2k - 3 \alpha$. In fact we can hope that there exists an extractor $E$ with $C(E(x,y)) = 2k - \alpha$ because $x$ and $y$ have each $k$ bits of randomness, of which they share $\alpha$ bits. For the stronger type of extraction in which we require that $E(x,y)$ has maximum possible Kolmogorov complexity even conditioned by any one of the input strings, we should aim for $C(E(x,y) \mid x) \approx  k - \alpha $ and  $C(E(x,y) \mid y) \approx k - \alpha$.

The latter optimal settings of parameters have been obtained for computable (but not  polynomial-time computable) Kolmogorov extractors by Zimand in~\cite{zim:c:kolmlimindep}, and in the stronger form in~\cite{zim:c:impossibamplific}.
\begin{theorem}[\cite{zim:c:impossibamplific}] Let $k(n)$ and $\alpha(n)$ be integers computable from $n$ such that $n \geq k(n) \geq \alpha(n) + 7 \log n + O(1)$. There exists a computable ensemble of functions $E: \zo^n \times \zo^n \mapping \zo^m$, where $m = k(n) - 7 \log n$ such that for all $(x,y) \in S_{k,\alpha}$, it holds that $C(E(x,y) \mid x) = m - \alpha(n) - O(1)$ and  $C(E(x,y) \mid y) = m - \alpha(n) - O(1)$.
\end{theorem}
\emph{Proof sketch.} As it is usually the case for constructions that achieve optimal parameters, we use the probabilistic method. The trick is to conceive the right type of balancing property that leads to the desired conclusion and that is satisfied by a random function. In our case, the balancing property, which we call \emph{rainbow balancing}, is somewhat complicated. 

\emph{Rainbow balanced tables}.  The novelty is that unlike the tables in Proposition~\ref{p:combextractor} and Proposition~\ref{p:combKolm}, where the balancing property refers to a single color per rectangle, now we require the balancing with respect to a different color for each column in the rectangle (and, separately, for each row). The table is of the form $E: [N] \times [N] \mapping [M]$. We fix a parameter $D$, which eventually will be taken to be $D \approx 2^{\alpha(n)}$. Let ${\cal A}_D$ be the collection of all sets of colors $A \subseteq [M]$, with size $|A| \approx M/D$. Let $B_1 \times B_2 \subseteq [N] \times [N]$ be a rectangle of size $K \times K$. We label the columns in $B_2$ as $B_2 =\{v_1 < v_2 < \ldots < v_K\}$. Let $\overline{A}= (A_1, A_2, \ldots, A_K)$ be a $K$-tuple with each $A_i \in {\cal A}_D$. In other words, for each column $v_i$ we fix a set of colors $A_i$. We say that a cell $(u,v_i)$ in $B_1 \times B_2$ is properly colored with respect to $B_2$ and $\overline{A}$ if $E(u,v_i) \in A_i$. Since $A_i \subseteq [M]$ and $|A_i| \approx M/D$, if $E$ is random, we expect the fraction of cells that are properly colored with respect to $B_2$ and $\overline{A}$ to be $\approx 1/D$. Similarly, we define the notion of a properly colored cell with respect to $B_1$ and a $K$-tuple  $\overline{A'}= (A'_1, A'_2, \ldots, A'_K)$. Finally, we say that the $[N]$-by-$[N]$ table $E$ colored with colors from$[M]$  is $(K,D)$-rainbow balanced if for all rectangles $B_1 \times B_2$ of size $K \times K$, for all $K$-tuples $\overline{A} \in ({\cal A}_D)^K$ and $\overline{A'} \in ({\cal A}_D)^K$, the fraction of cells in $B_1 \times B_2$ that are properly colored with respect to $B_2$ and $\overline{A}$ (and respectively, with respect to $B_1$ and $\overline{A'}$) is at most $2/D$.

\begin{table}
\begin{center}

{\tiny
\begin{tabular}{r|cc|ccccc|c}
&       &        & $B_2$   &         &         &          &         &    \\
& $u_1$ & $u_2$ & $\spadesuit$ & $\clubsuit$ & $\heartsuit$ &  $\spadesuit$ & $\clubsuit$ & $u_N$ \\
 \hline 
 $u_1$ & & & & & & & & \\
 $u_2$ & & & & & & & & \\
 $\cdot$ & & & & & & & & \\
 \hline
 $B_1$\quad \quad & $\cdot$ & $\cdot$ &$\spadesuit$ & $\clubsuit$& $\heartsuit$& $\spadesuit$& $\heartsuit$ & $\cdot$\\

 $\cdot$ & $\cdot$ & $\cdot$  &$\clubsuit$ &$\spadesuit$ &$\clubsuit$ &$\heartsuit$  &$\clubsuit$ & $\cdot$ \\

 $\cdot$ & $\cdot$ & $\cdot$ & $\clubsuit$& $\heartsuit$ &$\spadesuit$ & $\heartsuit$& $\spadesuit$ & $\cdot$ \\
 
 $\cdot$ & $\cdot$ & $\cdot$  &$\clubsuit$ &$\spadesuit$ &$\heartsuit$ &$\heartsuit$  &$\spadesuit$ & $\cdot$ \\
 
 $\cdot$ & $\cdot$ & $\cdot$  &$\heartsuit$ &$\spadesuit$ &$\clubsuit$ &$\heartsuit$  &$\clubsuit$ & $\cdot$ \\
 
 \hline
 $\cdot$  & & &  & & & & & \\
 $u_N$ & & & & & & & &
 \end{tabular}
 }
 \caption{\footnotesize {Rainbow-balanced table. For each column $v$ in $B_2$, we choose a set of colors $A_v \subseteq [M]$ of size $\approx M/D$, and we require that it does not appear more than a fraction of $2/D$ times in $B_1 \times \{v\}$. This should hold for all rectangles of size $K \times K$ and for all choices of $A_v$, and also if we switch the roles of columns and rows.}}
 
 \end{center}
 \end{table}
 A standard probabilistic analysis shows that a random table $E: [N] \times [N] \mapping [M]$ is $(K,D)$-rainbow balanced, provided $M < K$ and $D < K$ (in the latter inequalities we have omitted some small factors). 
 
We next present the construction. For readability, we hide some annoying small factors and therefore some of the parameters in our presentation are slightly imprecise. 
We take $d = \alpha(n) + c \log n$, for a constant $c$ that will be fixed later, $D= 2^d$, and $K = 2^{k(n)}$ (with a more careful analysis, we can take $d = \alpha(n) + O(1)$). The probabilistic argument shows that there exists a $(K,D)$ rainbow balanced table.
By brute-force we can effectively construct a $(K,D)$-rainbow balanced table $E$ for every $n$. Fix $(x,y) \in S_{k(n), \alpha(n)}$ and let $z = E(x,y)$. For the sake of contradiction suppose that $C(z \mid y) < m - d$. For each $v$, let $A_v = \{w \in [M] \mid C(w \mid v) < m - d\}$. It holds that $A_v \in {\cal A}_D$
 for all $v$. Let $B_x = \{u \in [N] \mid C(u \mid n) \leq C(x \mid n)\}$. Let us call a column $v$ \emph{bad} if the fraction of cells in $B_x \times \{v\}$ that are $A_v$-colored is larger than $2 \cdot (1/D)$. The number of bad columns is less than $K$, since otherwise the rainbow balancing property of $E$ would be violated. We infer that if $v$ is a bad column, then $C(v) \leq k(n)$. Since $C(y) \geq k(n)$, it follows that $y$ is a good column. Therefore the fraction of cells in the $B_x \times \{y\}$ strip of the table $E$ that have a color in $A_y$ is at most $2\cdot (1/D)$. Since $(x,y)$ is one of these cells, it follows that, given $y$, $x$ can be described by the rank $r$ of $(x,y)$ in an enumeration of the $A_y$-colored cells in the strip $B_x \times \{y\}$, a description of the table $E$, and by $O(\log n)$ additional bits necessary for doing the enumeration. Since $y$ is a good column, there are at most $2 \cdot (1/D) \cdot |B_x| \approx 2^{-d+1} \cdot 2^{C(x)}$ cells in $B_x \times \{y\}$ that are $A_y$-colored and, therefore, $\log r \leq C(x) - d + 1$. From here we obtain that $C(x \mid y) \leq C(x) - d + 1 + O(\log n) = C(x) - \alpha(n) - c \log n +O(\log n)$. Since $C(x \mid y) \geq C(x) - \alpha(n)$, we obtain a contradiction for an appropriate choice of the constant $c$. Consequently $C(z \mid y) \geq m - d = m - \alpha(n) - c \log n$. Similarly, $C(z \mid x) \geq m - \alpha(n) - c \log n$. With a more careful analysis the $c \log n$ term can be replaced with $O(1)$. Thus we have extracted $m \approx k(n)$ bits that have Kolmogorov complexity $\approx m - \alpha(n)$ conditioned by $x$ and also conditioned by $y$.~\qed
 
The proof of Theorem~\ref{t:fortnowKolmextract} is also based on the equivalence beteween multi-source extractors and Kolmogorov extractors.

\emph{Proof sketch of Theorem~\ref{t:fortnowKolmextract}.} The main tool is the polynomial-time multi-source extractor of Barak, Impagliazzo and Wigderson~\cite{bar-imp-wig:c:multisourceext}, which, for any $\sigma > 0$ and $c > 1$, uses $\ell = \poly(1/\sigma, c)$ independent sources of length $n$, with min-entropy $\sigma n$, and outputs a string of length $n$ that is $2^{-cn}$-close to $U_n$. Recall that the extractor in Theorem~\ref{t:fortnowKolmextract} works with a single source $x$ with randomness rate at least $\sigma$. The string $x$ is split into $\ell$ blocks $x_1, x_2, \ldots, x_\ell$, each of length $n$, with the intention of considering each block as a source. The main issue is that no independence property is guaranteed for the blocks $x_1, \ldots, x_\ell$ and therefore the extractor $E$ from~\cite{bar-imp-wig:c:multisourceext} cannot be used directly. However, one of the following cases must hold: (1) There exists $x_j$ with $C(x_j)$ low; in this case, since ${\rm rate}(x) \geq \sigma$, there must also exist  $x_i$ with ${\rm rate}(x_i) \geq \sigma + \gamma$, for some appropriate $\gamma$; (2) the dependency of $x_1, \ldots, x_\ell$ is high (\ie, the number of ``shared'' random bits is high); in this case again one can argue that there exists $x_i$ with ${\rm rate}(x_i) \geq \sigma + \gamma$; (3) the 
dependency of $x_1, \ldots, x_\ell$ is low; in this case, similarly to Theorem~\ref{t:ExtKolmExt}, the multi-source extractor $E$ is also a Kolmogorov extractor (with $\ell$ sources) and ${\rm rate}(E(x_1, \ldots, x_\ell))$ is close to $1$. Thus, either $x_i$, in cases $1$ and $2$, or $E(x_1, \ldots, x_\ell)$, in case $3$, has randomness rate higher than $x$. Iterating the procedure a constant number of times, we obtain a string with rate $1-\epsilon$. For this to work, we need to know, for each iteration, which one of Cases $1$, $2$, or $3$ holds and the index $i$ (for Cases $1$ and $2$). This constant information is given by the advice string $\alpha_x$.~\qed
\section{The infinite case}
\label{s:infinite}
 Effective Hausdorff dimension is the standard concept that quantifies the amount of randomness in an infinite binary sequence. This concept is obtained by an effectivization of the (classical) Hausdorff dimension, and, as we discuss in Section~\ref{s:HausKolm}, has an equivalent formulation in terms of the Kolmogorov complexity of the sequence prefixes. Namely, for each $x \in \zo^\infty$, ${\rm dim}(x) =    \lim \inf \frac{K(x \rest n)}{n} = \lim \inf\frac{C(x \rest n)}{n}$ (see Section~\ref{s:HausKolm} for the first equality; the second equality holds simply because the plain and the prefix Kolmogorov complexities are within $O(\log n)$ of each other).
 
 The issue of extraction from one infinite sequence has been first raised by  Reimann and Terwijn in 2003 (see~\cite{rei:t:thesis}). They asked whether for any sequence $x$ with ${\rm dim}(x) = 1/2$ there exists an effective transformation  $f$ such that ${\rm dim}(f(x)) > 1/2$ (the value $1/2$ is arbitrary; any positive rational number plays the same role). Formally, we identify an infinite sequence $x$ with the set of strings having $x$ as its characteristic sequence, $f$ is a Turing reduction corresponding to some oracle machine $M$, and $f(x)$ is the set computed by $M^x$, \ie, the $n$-th bit of $f(x)$ is $1$ iff $M^x$ accepts the $n$-th string in the lexicographical ordering of $\zo^*$. In case $M^x$ halts on every input, we also say that $f(x)$ is computed from $x$. Initially, some partial negative results have been obtained for transformations $f$ with certain restrictions.  Reimann and Terwijn~\cite{rei:t:thesis}   have shown that the answer is NO if we require that $f$ is a many-one reduction.  This result has been extended by Nies and Reimann~\cite{nie-rei:c:wtt-Kolm-increase} to wtt-reductions.
Bienvenu, Doty, and Stephan~\cite{bie-dot-ste:j:haussdimension}  have obtained an impossibility result for the general case of Turing reductions, which, however, is valid only for \emph{uniform} transformations. More precisely, building on the result of Nies and Reimann, they have shown that for all constants $c_1$ and $c_2$, with $0 < c_1 < c_2 < 1$, no single effective transformation is able to raise the dimension from $c_1$ to $c_2$ for all sequences with dimension at least $c_1$. Finally,  Miller~\cite{mill:j:KolmExtract} has fully solved the original question, by constructing a sequence $x$ with ${\rm dim}(x) = 1/2$ such that, for any Turing reduction $f$, ${\rm dim}(f(x)) \leq 1/2$ (or $f(x)$ does not exist). We present Miller's result in Section~\ref{s:miller}.

\subsection{Hausdorff dimension, effective Hausdorff dimension, and Kolmogorov complexity}
\label{s:HausKolm}
The Hausdorff dimension is a measure-theoretical tool used to create a distinction between sets that are too small to be differentiated by the usual Lebesgue measure (see, for example, Terry Tao's blog entry~\cite{tao:m:hausdorffdim} for an illuminating discussion). The sets of interest for us are subsets of $[0, 1]$ and we restrict the definitions to this case.

For $\sigma \in \zo^*$, $[\sigma]$ ($[\sigma]^{<\omega}$) is the set of all binary sequences (respectively, strings) having $\sigma$ as a prefix. For $V \subseteq \zo^*$, $[V] = \bigcup_{\sigma \in V} [\sigma]$ and $[V]^{< \omega} = \bigcup_{\sigma \in V} [\sigma]^{< \omega}$. If a string $\tau$ is a prefix of a string $\sigma$, we write $\tau \preceq \sigma$.
We use $\mu$ to denote the Lebesgue measure on $\zo^\infty$, determined by
$\mu([\sigma])=2^{-|\sigma|}$.

Let $A \subseteq \zo^\infty$ and $W \subseteq \zo^*$. $W$ is an $n$-cover of $A$ if all strings $\sigma$ in $W$ have $|\sigma| \geq n$  and $A \subseteq \bigcup_{\sigma \in W} [\sigma]$. For $s \in \real^{\geq 0}$, we define
$H^s_n(A) = \inf \{\sum_{\sigma \in W} 2^{-s|\sigma|} \mid W \mbox{ $n$-cover of $A$}\}$ and $H^s(A) = \lim_{n \rightarrow \infty} H^s_n(A)$ (the limit exists). It can be shown that there exists a unique $s$ such that for all $t > s$, $H^t(A)=0$ and for all $0 \leq u < s$, $H^u(A) = \infty$. The (classical) Hausdorff dimension of $A$ is defined as ${\rm dim}_H(A) = \inf \{ s \geq 0 \mid H^s(A) = 0 \}$. 

We illustrate the definitions with the analysis of the Cantor set. The underlying alphabet for this standard example is $\{0,1,2\}$, and we need to adapt the definitions for this setting by substituting ``$2^{-s|\sigma|}$'' with ``$3^{-s|\sigma|}$.'' The Cantor set is ${\cal C} = \{x \in \{0,1,2\}^\infty \mid x \mbox{ contains only 0s and 2 }\}$. The set $W= \{0,2\}^n$ is an $n$-cover of ${\cal C}$. If $s > 1/\log 3$, $H_n^s({\cal C}) \leq  \sum_{x \in W} 3^{-sn} = 2^n \cdot  3^{-sn}$, which goes to $0$ as $n$ grows. Thus, ${\rm dim}_H({\cal C}) \leq 1/\log 3$. If $s < 1/\log 3$, one can check that $W$ is an $n$-cover which yields the infimum in the definition of $H_n^s({\cal C})$. Thus, in this case, $H_n^s({\cal C}) = \sum_{x \in W} 3^{-sn} = 2^n \cdot  3^{-sn}$, which goes to infinity as $n$ grows. So, ${\rm dim}_H({\cal C}) \geq 1/\log 3$.  We conclude that the Hausdorff dimension of the Cantor set is $1/\log 3$.

We return to the binary alphabet. It can be shown that $H^s(A) = 0$ iff $\exists C \subseteq \zo^*$ such that $\sum_{\sigma \in C}2^{-s |\sigma|} < \infty$ and, for all $x \in A$, $\exists^\infty \sigma \in C$ with $x \in [\sigma]$. One way to define \emph{effective} Hausdorff dimension is to require that the set $C$ from above is computably enumerable.
\begin{definition}
\label{d:effHausdorff}
\begin{enumerate}
	\item[(1)] A set $A \subseteq \zo^\infty$ has effective $s$-dimensional Hausdorff dimension $0$ (where $s$ is a nonnegative rational number)  if $\exists C \subseteq \zo^*$, c.e., such that  $\sum_{\sigma \in C}2^{-s |\sigma|} < \infty$ and, for all $x \in A$, $\exists^\infty \sigma \in C$ with $x \in [\sigma]$. In this case we write $H_1^s(A) = 0$.
	\item[(2)] The effective Hausdorff dimension of $A \subseteq \zo^\infty$ is ${\rm dim}^1_H(A) = \inf \{s \geq 0 \mid H_1^s(A) = 0 \}$. If $A \in \zo^\infty$, instead of ${\rm dim}^1_H(\{A\})$, we simply write ${\rm dim}^1_H(A)$.
\end{enumerate}

\end{definition}
One can define effective Hausdorff dimension in a slightly different manner using Solovay tests. A Solovay $s$-test is given by a c.e. set $T$ of rational subintervals of $[0,1]$ such that $\sum_{I \in T} |I|^s < \infty$. A set $A \subseteq \zo^\infty$ is covered by $T$ if any element of $A$ is contained in infinitely many intervals of $T$. Reimann~\cite{rei:t:thesis} has shown that 
${\rm dim}^1_H(A) = \inf \{s \geq 0 \mid \mbox{ $A$ is covered by a Solovay $s$-test}\}$.
Since from now on we will be only using effective Hausdorff dimension, we abbreviate ${\rm dim}^1_H$ by just ${\rm dim}$.

For us, the most relevant is an alternate characterization of effective Hausdorff dimension based on Kolmogorov complexity.
\begin{theorem}[\cite{rya:j:dimension,may:j:dimension-kol,sta:j:dimension}] 
\label{t:HausdorffKolm}
For any $x \in \zo^\infty$, ${\rm dim}(x) = \lim \inf \frac{K(x \rest n)}{n}$.
\end{theorem}
$\proof$ ``${\rm dim}(x) \leq \lim \inf \frac{K(x \rest n)}{n}$." Let $s > \lim \inf \frac{K(x \rest n)}{n}$. We show that $H_1^s(x) = 0$, which implies ${\rm dim}(x) < s$, from which the conclusion follows. Take $C = \{\sigma \in \zo^* \mid K(\sigma) < s |\sigma|\}$. Note that: (a) $C$ is c.e., (b) $(\exists^\infty n) ~x \rest n \in C$, (c) $\sum_{\sigma \in C} 2^{-s|\sigma|} < \infty$ (because $\sigma \in C$ implies $K(\sigma) < s |\sigma|$ and therefore $2^{-K(\sigma)} > 2^{-s|\sigma|}$ and $\sum 2^{-K(\sigma)} \leq 1$ by Kraft-Chaitin inequality). Thus (1)~in Definition~\ref{d:effHausdorff} is satisfied.

``$\lim \inf \frac{K(x \rest n)}{n} \leq {\rm dim}(x)$." Let $s$ be such that $H^s_1(x) = 0$. We show that $\lim \inf \frac{K(x \rest n)}{n} \leq s$. We know that there exists a c.e. set $C$ such that $\sum_{\sigma \in C} 2^{-s |\sigma|} < \infty$ and $\exists^\infty \sigma \in C$ with $x \in [\sigma]$. For some constant $c$, $\sum_{\sigma \in C} 2^{-s |\sigma|-c} \leq 1$. Using the other direction of Kraft-Chaitin theorem, it follows that for all $\sigma \in C$, $K(\sigma) \leq s |\sigma| +O(1)$ and therefore $\frac{K(\sigma) - O(1)}{|\sigma|} \leq s$. Consequently, $\exists^\infty n$ $\frac{K(x \rest n) - O(1)}{n} \leq s$ which implies $\lim \inf \frac{K(x \rest n)}{n} \leq s$.~\qed

\subsection{A strong impossibility result: Miller's theorem}
\label{s:miller}
\begin{theorem}[\cite{mill:j:KolmExtract}]
\label{t:miller}
There exists $A \in \zo^\infty$, with ${\rm dim}(A) = 1/2$, such that any $B \in \zo^\infty$ computable from $A$ has ${\rm dim}(B) \leq 1/2$.
\end{theorem}
$\proof$ We use the notation introduced in Section~\ref{s:HausKolm}.
For $S \subseteq \zo^*$, we define the direct weight of $S$ by $\dw(S) = \sum_{\sigma \in S} 2^{-|\sigma|/2}$ and the weight of $S$ by $\w(S) = \inf \{ \dw(V) \mid [S] \subseteq [V]\}$. A set $V$ that achieves the infimum in the definition of $\w(S)$ is called an optimal cover of $S$. An optimal cover exists for any set $S$ for the following reasons. If $S$ is finite, then it is not optimal to consider in a cover of $[S]$ a string $\tau$ that does not have an extension in $S$; so there are only finitely many candidates for an optimal cover. If $S \subseteq \zo^*$ is an infinite set, then consider an enumeration $\{S_t\}_{t \in \nat}$ of $S$, \ie, an increasing sequence of finite sets such that $S = \cup S_t$. Let $S_t^{oc}$ be the optimal cover of $S_t$. The only way for a string $\sigma \in S_t^{oc}$ to not be in $S_{t+1}^{oc}$ is if there exists a string $\tau \preceq \sigma$ in $S_{t+1}^{oc}$. This shows that $[S_t^{oc}] \subseteq [S_{t+1}^{oc}]$ and that the sets $S_t^{oc}$ have a limit $V$, with $[S] \subseteq [V]$. This set has the property that
$\dw(V) = \w(S)$. So we define $S^{oc}(S)$ to be the set $V$ with $[S] \subseteq [V]$ and $\dw(V) = \w(S)$ (if there is a tie, we pick $V$ with the minimum measure). If $S$ is \ce, it does not follow that $S^{oc}$ is c.e.  However, if $\{S_t\}_{t \in \nat}$ is an effective enumeration of $S$ and $V = \cup S_t^{oc}$, then $V$ is \ce, $[V] = [S^{oc}]$ and for any prefix-free set $P \subseteq V$ it holds that $\dw(P) \leq \dw(S^{oc}) = \w(S)$.
A key fact is shown in the next lemma: For any \ce\ $S$, the measure of $[S^{oc}]$  (viewed as an infinite binary sequence obtained through binary expansion)  has effective dimension at most $1/2$.
\begin{lemma}
\label{l:onehalf}
If $S$ is \ce, then $\dim(\mu([S^{oc}])) \leq 1/2$.
\end{lemma}
$\proof$
If $S^{oc}$ is finite, then $\mu([S^{oc}])$ is rational and thus has effective dimension $0$. So assume that $S^{oc}$ is infinite. Let $w = \w(S)$ and let $V$ be the set from the paragraph preceding the lemma. Let $(V_t)_{t \in \nat}$ be an effective enumeration of $V$, with $V_0 = \emptyset$. For an arbitrary $s > 1/2$, we construct a Solovay $s$-test $T$ that covers $\mu([V])$. Since $[V] = [S^{oc}]$, this will establish the lemma. $T$ has two parts, $T_0$ and $T_1$.
\begin{enumerate}
	\item [(a)] If $\tau \in V_{t+1} - V_t$, then put $[\mu([V_{t+1}]), \mu([V_{t+1}]) + 2^{-|\tau|}]$ into $T_0$.
	\item [(b)] If, for some $k, n \in \nat$, $\mu([V_t \cap \zo^{>n}]) \leq k \cdot 2^{-n}$  and $\mu([V_{t+1} \cap \zo^{>n}]) > k \cdot 2^{-n}$,  then put $[\mu([V_{t+1}]), \mu([V_{t+1}]) + 2^{-n}]$ into $T_1$.
\end{enumerate}
Clearly, $T = T_0 \cup T_1$ is a \ce\ set of rational intervals. Let us show that $T$ is a Solovay $s$-test. First we analyze $T_0$.
\[
\begin{array}{l}
\sum_{I \in T_0} |I|^s = \sum_{\tau \in V} 2^{-s|\tau|} = \sum_n 2^{-sn} |V \cap \zo^n| \\
= \sum_n 2^{(1/2-s)n} 2^{-n/2} \cdot |V \cap \zo^n| = \sum_n 2^{(1/2-s)n} \cdot \dw(V \cap \zo^n) \\
\leq \sum_n 2^{(1/2-s)n} \cdot w = w \cdot \sum_n 2^{(1/2-s)n} < \infty,
\end{array}
\]
where in the transition to the last line we have used that $V \cap \zo^n$ is prefix-free and the above property of $V$.
We move to $T_1$. Fix $n$ and let $k$ be the number of intervals of length $2^{-n}$ added to $T_1$. By construction, $k \cdot 2^{-n} < \mu([V \cap \zo^{>n}])$. Let $P \subseteq V \cap \zo^{>n}$ be a prefix-free set such that $[P] = [V \cap \zo^{>n}]$. Then $\mu([P]) = \sum_{\tau \in P} 2^{-|\tau|} < \sum_{\tau \in P} 2^{-|\tau|/2 - n/2} = 2^{-n/2} \sum_{\tau \in P} 2^{-|\tau|/2} = 2^{-n/2}\cdot \dw(P)$. So, $k \cdot 2^{-n} < 2^{-n/2}\cdot \dw(P) \leq 2^{-n/2}\cdot w$ and thus $k < 2^{n/2} \cdot w$. Therefore, $\sum_{I \in T_1} |I|^s < \sum_n 2^{n/2} \cdot w \cdot (2^{-n})^s < \infty$. We conclude that $T_1$ is a Solovay $s$-test, and so $T$ is a Solovay $s$-test. Next, we show that $T$ covers $\mu([V])$. Call $\tau \in V$ timely if only strings longer than $\tau$ enter $V$ after $\tau$. Let us fix a timely $\tau$, let $n = |\tau|$ and let $t+1$ be the stage when $\tau$ enters $V$. We claim that there is an interval of length $2^{-n}$ in $T$ that contains $\mu([V])$. When $\tau$ enters $V$, we put the interval $[\mu([V_{t+1}]), \mu([V_{t+1}] + 2^{-n}]$ in $T_0$. Let $I = 
[\mu([V_{u}]), \mu([V_{u}] + 2^{-n}]$ be the last interval of length $2^{-n}$ added to $T$. If $\mu([V]) \not \in I$, then $\mu([V]) > \mu([V_u]) + 2^{-n}$. By the construction of $T_1$, another interval of length $2^{-n}$ is added to $T_1 \subseteq T$ after stage $u$, which is a contradiction. Thus, we conclude that for every $n$ that is the length of a timely element of $V$, there is an interval of length $2^{-n}$ in $T$ that contains $\mu([V])$. Since there are infinitely many timely elements, $\mu([V])$ is covered by $T$.~\qed

\emph{Construction of set $A$.}  Let $(\Psi_e)_{e \in \nat}$ be an effective enumeration of all oracle Turing machines, and let $\Psi_e^A \rest k$ denote the initial segment of length $k$ of the characteristic sequence of the set accepted by $\Psi_e$ with oracle $A$. 

The set $A$ is constructed in stages so that it satisfies all requirements $R_{e,n}$ defined as
\smallskip

$R_{e,n}$: If $\Psi_e^A$ is total, then $(\exists k > n) K(\Psi_e^A \rest k) \leq (1/2+ 2\cdot 2^{-n})k$,
\smallskip

which implies that any set computed from $A$ has effective dimension at most $1/2$.
The construction defines a sequence of \emph{conditions}. A condition is a pair $\langle\sigma, S\rangle$, where $\sigma \in \zo^*$, $S \subseteq [\sigma]^{< \omega}$ is a \ce\ set, and $\sigma \not \in S^{oc}$. The string $\sigma$ will be an initial segment of $A$, and the set $S$ defines some obstructions for $A$ in the sense that we need to guarantee that $A \not \in [S^{oc}]$. Thus, we define $P_{\langle\sigma, S\rangle} = [\sigma] - [S^{oc}]$, which is viewed as the set of possibilities for $A$ according to condition $\langle\sigma, S\rangle$.

At stage $t$, we define the condition $(\sigma_t, S_t)$  so that any set $A$ in $P_{\langle \sigma_t, S_t \rangle}$ satisfies requirement $R_{e,n}$ for $\langle e,n\rangle = t$.  We make sure that  $P_{\langle\sigma_{t+1}, S_{t+1}\rangle} \subseteq P_{\langle\sigma_t, S_t\rangle}$ and that all $P_{\langle\sigma_t, S_t\rangle}$ are not empty. Finally, we take $A$ to be the limit of the strings $\sigma_t$ and we also ensure that $\dim(A) = 1/2$.

We first list a few useful properties of  conditions $\langle\sigma, S\rangle$.

\emph{Fact 1.} If $P_{\langle\sigma, S\rangle}$ is not empty, then it has positive measure. This follows from a calculation similar to one used in Lemma~\ref{l:onehalf}.

\emph{Fact 2.} $\dim(\mu(P_{\langle\sigma, S\rangle})) \leq 1/2$. Note that $\mu(P_{\langle\sigma,S\rangle}) = 2^{-|\sigma|} - \mu ([S^{oc} \cap [\sigma]^{< \omega}])$ and we invoke Lemma~\ref{l:onehalf}.

\emph{Fact 3.} If $\langle\sigma_1, S_1\rangle, \ldots, \langle\sigma_n , S_n\rangle$ are conditions such that $P_{\langle\sigma_1, S_1\rangle} \cap \ldots \cap P_{\langle\sigma_n, S_n\rangle}$ has positive measure, then there exists a condition $\langle\tau, T\rangle$ such that
$P_{\langle\tau, T\rangle} \subseteq P_{\langle\sigma_i, S_i\rangle}$, for all $1 \leq i \leq n$.

Armed with these facts, we proceed to describe the construction. At stage $t=0$, we take the condition $\langle\lambda, S_0\rangle$, where $S_0 = \{ \sigma \in \zo^* \mid K(\sigma) \leq |\sigma|/2\}$. It can be checked that $\langle\lambda, S_0\rangle$ is indeed a condition. Since $A \in P_{\langle\lambda, S_0\rangle}$, it follows that $\dim(A) \geq 1/2$.  Since $A$ is of course computable from $A$, the construction also guarantees that $\dim(A) \leq 1/2$, and thus $\dim(A) = 1/2$.

At stage $t+1 = \langle e,n\rangle$ we satisfy the requierement $R_{e,n}$. First we choose $b \in \nat$ such that $\mu(P_{\langle\sigma_t, S_t\rangle}) > 2^{-b}$. Let $\sigma \in \zo^*$ be an initial segment of the binary expansion of $\mu(P_{\langle\sigma_t, S_t\rangle})$ of length $m > n + b$, where $m$ is sufficiently large for what follows, and $K(\sigma) \leq (1/2 + 2^{-n})(m-b)$. Such a string $\sigma$ exists because $\dim(\mu(P_{\langle\sigma_t, S_t\rangle})) \leq 1/2$.
For each $\tau \in \zo^*$, we define $T_\tau = \{\nu \succ \sigma_t, \tau \preceq \Psi_e^{\nu}\}$. There are two cases to consider, depending on whether there exists or not $\tau \in \zo^{m-b}$ such that $S_t \cup T_\tau$ has large measure (specifically, larger than $2^{-|\sigma_t|} - .\sigma$).

\emph{Case 1.} There exists $\tau \in \zo^{m-b}$ such that $\mu(P_{\langle\sigma_t, S_t \cup T_\tau\rangle}) < .\sigma$ (\ie, $\mu(S_t \cup T_\tau)$ is large, that is there are many extensions of $\sigma_t$ that compute via $\Psi_e$ the same initial segment $\tau$).  Note that, given $\sigma$ and $t$, one can enumerate the strings satisfying the above property. Let $\tau$ be the first such string in the enumeration. Since $\tau$ is essentially described by $\sigma$ and a few additional bits, it follows that $K(\tau) \leq K(\sigma) + 2^{-n}(m-b)$ (if $m$ is sufficiently large), and thus $K(\tau) \leq (1/2 + 2 \cdot 2^{-n})(m-b)$. On the other hand, since $\mu(P_{\langle\sigma_t, S_t \cup T_\tau\rangle}) < .\sigma$ and $\mu(P_{\langle\sigma_t, S_t\rangle}) \geq .\sigma$, it follows that there exists a string $\sigma_{t+1} \in T_\tau$ such that $\sigma_{t+1} \not\subseteq [S_t^{oc}]$. We take $S_{t+1} = [\sigma_{t+1}]^{<\omega} \cap S_t$. It can be checked that $\langle\sigma_{t+1}, S_{t+1}\rangle$ is a valid condition with $\emptyset \not= P_{\langle\sigma_{t+1}, S_{t+1}\rangle} \subseteq P_{\langle\sigma_{t}, S_{t}\rangle}$. It remains to check that $R_{e,n}$ has been satisfied. By construction, $\sigma_{t+1} \preceq A$ and $\tau \preceq \Psi_e^{\sigma_{t+1}}$ (because $\sigma_{t+1} \in T_\tau$). Thus, $\tau \preceq \Psi_e^{A}$. Also, $|\tau| = m-b > n$. Then, $K(\Psi_e^A \mid m-b) = K(\tau) \leq (1/2 + 2 \cdot 2^{-n})(m-b)$.

\emph{Case 2.} There is no $\tau$ as in Case 1. We satisfy $R_{e,n}$, by guaranteeing that $\psi_e^A$ is not total. In Case 2, $\mu(P_{\langle\sigma_t, S_t \cup T_\tau\rangle}) \geq .\sigma$ for all $\tau \in \zo^{m-b}$. Since $\mu(P_{\langle\sigma_t, S_t\rangle}) < .\sigma + 2^{-m}$, it follows that $\mu(P_{\langle\sigma_t, S_t\rangle} - P_{\langle\sigma_t, S_t \cup T_\tau\rangle}) < 2^{-m}$, \ie, the obstructions added by each $T_\tau$ have very small measure. There are $2^{m-b}$ such $T_\tau$ and thus the union of obstructions added by all $T_\tau$ has measure $\leq 2^{m-b}\cdot 2^{-m} = 2^{-b}$. Since $P_{\langle\sigma_t, S_t\rangle}$ has measure $>2^{-b}$, it follows that the measure $\bigcap_{\tau \in \zo^{m-b}} P_{\langle\sigma_t, S_t \cup T_\tau\rangle}$ has positive measure. Thus, by Fact 3, there exists a condition $\langle\sigma_{t+1}, S_{t+1}\rangle$ that extends $\langle\sigma_t, S_t \cup T_\tau\rangle$ for all $\tau \in \zo^{m-b}$. Now, suppose that $\Psi_e^A$ is total and let $\tau  = \Psi_e^A \rest (m-b)$. Since $\sigma_t \prec A$, there is some $\rho \prec A$ in $T_\tau$, which implies that $A \in [S_t \cup T_\tau] \subseteq [(S_t \cup T_\tau)^{oc}]$ and hence $A \not \in P_{\langle\sigma_t, S_t \cup T_\tau\rangle}$, contradiction.~\qed

\subsection{Positive results regarding Kolmogorov extraction from infinite sequences}
Taking into account Miller's Theorem~\ref{t:miller}, one can hope for positive results only if

(a) the Kolmogorov extractor uses at least two independent sequences, or

(b) it uses one sequence but the randomness requirement on the output is weaker than effective Hausdorff dimension $1$.

We present the main results for these two situations. There is no room here for proofs; self-contained proofs can be found in Chapter 12 of~\cite{dow-hir:b:algrandom}.

Regarding (a), a first observation is that it is not obvious what \emph{independence} means for sequences. 
Levin~\cite{lev:j:mutualinformation} has suggested a notion of algorithmical mutual information based on the  corresponding concept in classical information theory. However, Levin's proposal is technically complicated and some basic questions remain open. For example, in Levin's setting, it is not clear if every sequence is dependent with itself.
Finding the ``right'' definition of independence for sequences is an important open problem in algorithmical randomness theory (see~\cite{down:m:openq}). 
Calude and Zimand~\cite{cal-zim:j:algindep}  have several proposals that are perhaps not tight but are natural and good enough for the working mathematician. In particular, a notion of independence from~\cite{cal-zim:j:algindep}, which is called $C$-independence in~\cite{dow-hir:b:algrandom}, is sufficient for Kolmogorov extraction. We say that sequences $x$ and $y$ are $C$-independent if $C(x\rest n ~y \rest m) \geq C(x \rest n) + C(y \rest m) - O(\log n + \log m)$, for all $n$ and $m$. With this definition, Kolmogorov extraction is possible in situation (a).
\begin{theorem}[\cite{zim:j:extractKolm}] 
\label{t:twosources}
For every rational number $\sigma > 0$, there exists a Turing reduction (actually a truth-table reduction)  $f$, such that for all $C$-independent sequences $x$ and $y$, with ${\rm dim}(x) \geq \sigma$ and ${\rm dim}(y) \geq \sigma$, it holds that ${\rm dim}(f(x,y)) = 1$. Moreover, $f$ is uniform in $\sigma$.
\end{theorem}
For (b), the relaxation is to require that the \emph{effective packing dimension} of the output is close to $1$. The effective packing dimension of a sequence $x$, denoted ${\rm Dim}(x)$, is in many ways the dual of the effective Hausdorff dimension ${\rm dim}(x)$, and, analogously to Theorem~\ref{t:HausdorffKolm}, admits a characterization based on Kolmogorov complexity: ${\rm Dim}(x) = \lim \sup \frac{C(x \rest n)}{n}$.
Fortnow et al.~\cite{fhpvw:c:extractKol} show that it is possible to construct a sequence with packing dimension close to $1$ from any sequence $x$ with ${\rm Dim}(x) > 0$ and a lower bound of ${\rm Dim}(x)$.
\begin{theorem}[\cite{fhpvw:c:extractKol}] 
\label{t:fhpvwtwo}
For every $\epsilon > 0$ and every $\sigma >0$, there exists a Turing reduction $f$ such that for every sequence $x$ with ${\rm Dim}(x) \geq \sigma$, it holds that ${\rm Dim}(f(x)) \geq 1 - \epsilon$. Moreover, $f$ is a polynomial-time computable reduction.
\end{theorem}

Conidis~\cite{con:m:packdim} shows that $1-\epsilon$ cannot be replaced by $1$ in Theorem~\ref{t:fhpvwtwo}. His result,  which can be viewed as the analog of Miller's Theorem for effective packing dimension, shows the existence of a sequence $x$ with ${\rm Dim}(x) \geq 1/4$ such that for every Turing reduction $f$, ${\rm Dim}(f(x)) < 1$ (or $f(x)$ is not defined). On the other hand, it is open whether from a sequence $x$ with ${\rm dim}(x) > 0$ it is possible to effectively construct $f(x)$ with ${\rm Dim}(f(x)) = 1$.

Doty~\cite{dot:j:dimextractors} shows that from any sequence $x$ with ${\rm dim}(x) > 0$ and a good upper bound of ${\rm dim}(x)$, one can construct a sequence with effective packing dimension close to $1$.
\begin{theorem}[\cite{dot:j:dimextractors}] For every rational $\beta$ there exists a Turing reduction $f$ such that for every sequence $x$ with ${\rm dim}(x) < \beta$ it holds that ${\rm Dim}(f(x)) \geq 1 - \epsilon$, where
$\epsilon = (\beta/{\rm dim}(x)) - 1$.
\end{theorem}
Another related result is due to Bienvenu, Doty, and Stephan~\cite{bie-dot-ste:j:haussdimension}.
\begin{theorem}[\cite{bie-dot-ste:j:haussdimension}] For every $\epsilon > 0$, there exists a Turing reduction $f$ such that for every sequence $x$, it holds that ${\rm dim}(f(x)) \geq ({\rm dim}(x)/{\rm Dim}(x)) - \epsilon$. Thus, if ${\rm dim}(x) = {\rm Dim}(x)$, we have ${\rm dim}(f(x))= 1-\epsilon$.
\end{theorem}

\section{Applications}

We discuss here several applications of Kolmogorov extractors.
\smallskip

(a)\emph{Hitting properties.}
Many technical utilizations of extractors exploit the fact that an extractor $E$ maps its domain almost uniformly to its range and therefore $E$ ``hits" any subset of its range proportionally to the density of the set.  The Kolmogorov complexity spin allows the derivation of special properties regarding the way in which a Kolmogorov extractor hits computable subsets of its range.
For instance, let $A \subseteq \zo^*$ be a set such that $A^{=n}$ is computable by circuits of size $s(n)$. Then each string $z$ in $A^{=n}$ has complexity $C(z \mid n) \leq s(n) + \log |A^{=n}| + c$, for some constant $c$. Let $E$ be a Kolmogorov extractor such that for every $(x,y) \in S_{k,\alpha}$, $C(E(x,y) \mid n) > s(n) + \log |A^{=n}| + c$. Then we deduce that $E(S_{k,\alpha})$ does not hit $A$ at all, \ie, for all $(x,y) \in S_{k,\alpha}$, $E(x,y) \not \in A$.
\smallskip

The most natural domain where Kolmogorov extractors have applications is the Kolmogorov complexity theory. We discuss two examples from the papers~\cite{zim:c:countingstrings} and~\cite{zim:c:impossibamplific}.
\smallskip

(b)\emph{Counting dependent strings.}
Given an $n$-bit string $x$ and a natural number $\alpha$, it is useful  to estimate the number of $n$-bit strings $y$ about which $x$ has $\alpha$ bits of information, \ie, the size of $B_{x, \alpha} = \{y \in \zo^n \mid C(y \mid n) - C(y \mid x) \geq \alpha\}$. The upper bound $|B_{x,\alpha}| < c\cdot 2^{n-\alpha}$, for a constant $c$, is easy to derive. For finding a lower bound, there is a  ``normal'' and simple approach that is best illustrated when $x$ is random. In this case, the prefix $x(1:\alpha)$ of $x$ of length $\alpha$ is also random and, therefore, if we take $z$ to be an $(n-\alpha)$ long string that is random conditioned by $x(1:\alpha)$, then $C(z x(1:\alpha)) = n - O(\log n)$, $C(zx(1:\alpha) \mid x(1:\alpha)) = n - \alpha - O(\log n)$, and thus, $zx(1:\alpha) \in B_{x, \alpha + O(\log n)}$. There are approximately $2^{n - \alpha}$ strings $z$ as above, and this leads to a lower bound of $2^{n-\alpha}$ for $|B_{x, \alpha + O(\log n)}|$, which implies a lower bound of $(1/\poly(n))2^{n-\alpha}$ for $|B_{x,\alpha}|$. This method is so basic and natural that it looks hard to beat. However, using properties of Kolmogorov complexity extractors, we derive a better lower bound for $|B_{x,\alpha}|$ that does not have the slack of $1/\poly(n)$, in case $C(x) \geq \alpha + O(\log n)$ and $\alpha$ is computable from $n$ (even if $\alpha$ is not computable from $n$, the new method gives a tighter estimation than the above ``normal'' method).  
Recall that there exists an extractor $E$ that on input $(x,y) \in S_{k,\alpha}$  outputs an $m$-bit string $z$ with $m \approx k$ and Kolmogorov complexity equal to $m - \alpha - O(1)$ even conditioned by any one of the input strings. 
We fix $x \in \zo^n$ with $C(x) \geq k$. Let $z$ be the most popular image of the function $E$ restricted to $\{x\} \times \zo^n$.  Because it is distinguishable from all other strings, given $x$, $z$ can be described with only $O(1)$ bits.  Choosing $m$ just slightly larger than $\alpha$ we arrange that $C(z \mid x) < m-\alpha -O(1)$. This implies that all the preimages of $z$ under  $E$ restricted as above are are 
\emph{bad-for-extraction}, \ie, they are not in $S_{k,\alpha}$.
Since the size of $E^{-1}(z) \cap (\{x\}\ \times \zo^n)$ is at least $2^{n-m}$, we see that at least $2^{n-m}$ pairs $(x,y)$ are bad-for-extraction. A pair of strings $(x,y)$ is bad-for-extraction if either $y$ has Kolmogorov complexity below $k$ (and it is easy to find an upper bound on the number of such strings), or if $y \in B_{x, \alpha}$. This leads to the lower bound $|B_{x,\alpha}| \geq (1/C)2^{n-\alpha} - \poly(n)2^\alpha$.
\smallskip

(c) \emph{Impossibility of independence amplification.}
The dependency of two strings $x$ and $y$ is another attribute (besides randomness deficiency) of randomness imperfection. Therefore, one would like to decrease the dependency of strings (in other words, to amplify their independence), \ie, one would like to have computable functions $f_1$ and $f_2$ such that for all strings $x$ and $y$ satisfying certain properties, ${\rm dep}(f_1(x,y), f_2(x,y)) < {\rm dep}(x,y)$. Unfortunately, effective independence amplification is impossible for strings $(x,y) \in S_{k,\alpha}$ and this can be easily shown using Kolmogorov extractors. Indeed, if for all $(x,y) \in S_{k,\alpha}$, ${\rm dep}(f_1(x,y), f_2(x,y)) = \beta < \alpha - O(\log \alpha)$, then, from $f_1(x,y)$ and $f_2(x,y)$, one could effectively produce a string $z$ with randomness deficiency $\beta$, and this contradicts the ``curse of dependency'' Theorem~\ref{t:cursedep}.

\section{Acknowledgements}
I am grateful to Andrei Romashcenko for his very helpful comments.

\if01
\newcommand{\etalchar}[1]{$^{#1}$}

\fi
\bibliography{c:/book-text/theory}

\bibliographystyle{alpha}

\end{document}